\documentclass[journal]{IEEEtran}
\ifCLASSINFOpdf
\else
\fi

\usepackage{soul}
\usepackage[utf8]{inputenc} % allow utf-8 input
\usepackage[T1]{fontenc}    % use 8-bit T1 fonts
\usepackage{hyperref}       % hyperlinks
\usepackage{url}            % simple URL typesetting
\usepackage{booktabs}       % professional-quality tables
\usepackage{amsfonts}       % blackboard math symbols
\usepackage{nicefrac}       % compact symbols for 1/2, etc.
\usepackage{microtype}      % microtypography
\usepackage{lipsum}
\usepackage{graphicx}
\usepackage{array,multirow}
\usepackage{amsmath}
\usepackage{xcolor}
\graphicspath{ {./images/} }

\usepackage{makecell}
\usepackage[
singlelinecheck=false % <-- important
]{caption}

\usepackage{cite}  % converts [1],[2] to [1,2]

\makeatletter
\def\@citex[#1]#2{\leavevmode
\let\@citea\@empty
\@cite{\@for\@citeb:=#2\do
{\@citea\def\@citea{,\penalty\@m\ }%
\edef\@citeb{\expandafter\@firstofone\@citeb\@empty}%
\if@filesw\immediate\write\@auxout{\string\citation{\@citeb}}\fi
\@ifundefined{b@\@citeb}{\hbox{\reset@font\bfseries ?}%
\G@refundefinedtrue
\@latex@warning
{Citation `\@citeb' on page \thepage \space undefined}}%
{\@cite@ofmt{\csname b@\@citeb\endcsname}}}}{#1}}
\makeatother

\begin{document}
\title{EVALUATING KNOWLEDGE TRANSFER IN NEURAL NETWORK FOR MEDICAL IMAGES}

\author{Sina~Akbarian,
        Laleh~Seyyed-kalantari,
        Farzad~Khalvati,
        and~Elham~Dolatabadi% <-this % stops a space
\thanks{S. Akbarian is with Public Health Ontario, University of Toronto, and Vector Institute, 661 University, ON, M5G 1M1, Canada.}
\thanks{F. Khalvati is with University of Toronto.}

\thanks{L. Seyyed-kalantari and E. Dolatabadi are with University of Toronto, and Vector Institute, 661 University, ON, M5G 1M1, Canada. (\emph{e-mail: elham.dolatabadi@mail.utoronto.ca})}}

\maketitle

\begin{abstract}
Deep learning and knowledge transfer techniques have permeated the field of medical imaging and are considered as key approaches for revolutionizing diagnostic imaging practices. However, there are still challenges for the successful integration of deep learning into medical imaging tasks due to a lack of large annotated imaging data. To address this issue, we propose a teacher-student learning framework to transfer knowledge from a carefully pre-trained convolutional neural network (CNN) teacher to a student CNN. In this study, we explore the performance of knowledge transfer in the medical imaging setting. We investigate the proposed network’s performance when the student network is trained on a small dataset (target dataset) as well as when teacher’s and student’s domains are distinct. The performances of the CNN models are evaluated on three medical imaging datasets including Diabetic Retinopathy, CheXpert, and ChestX-ray8. Our results indicate that the teacher-student learning framework outperforms transfer learning for small imaging datasets. Particularly, the teacher-student learning framework improves the area under the ROC Curve (AUC) of the CNN model on a small sample of CheXpert (n=5\textit{k}) by 4\% and on ChestX-ray8 (n=5.6\textit{k}) by 9\%. In addition to small training data size, we also demonstrate a clear advantage of the teacher-student learning framework in the medical imaging setting compared to transfer learning. We observe that the teacher-student network holds a great promise not only to improve the performance of diagnosis but also to reduce overfitting when the dataset is small.

\end{abstract}


% \fi
\IEEEpeerreviewmaketitle

\section{Introduction}
Medical imaging is widely used for diagnosing several life-threatening diseases. However, shortage of expert human resources to read and interpret medical imaging exams puts patients’ lives at risk~\cite{rimmer_radiologist_2017,bastawrous_improving_2017}. Therefore, finding a reliable alternative for expediting reading and interpreting medical images is critical in order to improve diagnosis and consequently treatment of diseases~\cite{torres-mejia_radiographers_2015}. Recently, Artificial Intelligence (AI)-based systems especially state-of-the-art Deep Neural Network (DNN) models have proved to be effective in improving clinical decision making for medical imaging diagnostics~\cite{lakhani_deep_2017, rajpurkar_chexnet_2017,lundervold2019overview}. However, training a DNN from random initialization to achieve high accuracy is compute-intensive, memory-demanding, and generally requires a large amount of annotated data that is not always easy to collect in the medical domain. Knowledge transfer has gained much attention in the research community in order to address these shortcomings with training DNN models~\cite{hinton2015distilling,bucilu2006model,jang2019learning,gutstein2008knowledge,romero2014fitnets,ZagoruykoK16a}.\\

Knowledge transfer from a source domain to a target domain is a technique to facilitate the training process of the DNN on smaller datasets. Recently, several approaches on the knowledge transfer technique have been proposed to maintain the performance of DNN models while using small training datasets~\cite{Pan2009,knoll2019assessment,bucilu2006model,hinton2015distilling,Yim_2017}. One popular approach in knowledge transfer is transfer learning in which a model already pre-trained on a large source dataset (such as ImageNet~\cite{imagenet_cvpr09}) is fine-tuned on a target dataset (e.g. medical images) with minimal modifications where some of the parameters remain frozen during training~\cite{Pan2009}. A pre-trained network trained on large datasets with thousands of classes, various illumination conditions, different backgrounds, and orientation is a powerful tool to extract features~\cite{Power-pretrain} even in a very small and noisy target data regime. Using transfer learning, the network retains its ability to extract low-level features learned from the source domain, and learns how to combine them to detect complex patterns on the target domain~\cite{knoll2019assessment}. Transfer learning has been the basis for DNN-based medical imaging diagnosis such as skin cancer~\cite{esteva_dermatologist-level_2017,codella2017deep}, chest X-rays~\cite{ raghu_transfusion:_2019,irvin_chexpert:_2019, rajpurkar_deep_2018,wang_chestx-ray8:_2017, CheXclusion_2020,yao_learning_2017}, Diabetic Retinopathy~\cite{pratt2016convolutional,li2017convolutional,masood2017identification,lam2018automated,benson2018transfer}, Alzheimer’s Disease~\cite{Alzheimer2018,liu2014early}, and sleep monitoring~\cite{sina-osa}. However, in an empirical study conducted by Raghu et al.~\cite{raghu_transfusion:_2019}, it has been shown that using transfer learning from ImageNet to medical images, the parameters of the convolutional neural network (CNN) models do not update drastically during the fine-tuning. This study also showed that smaller architectures trained on medical image datasets from scratch can perform similar to the transfer learning from large models. Moreover, Jang {\em et al.}~\cite{Jang2019LearningWA} also reported that transfer learning may not help if the two tasks and/or datasets are semantically distinct.\\

Another popular approach in knowledge transfer is a teacher-student learning framework that has been actively studied in recent years in order to improve the transfer of knowledge for both in-domain and cross-domain tasks~\cite{hinton2015distilling,Yim_2017,romero2014fitnets,ZagoruykoK16a}. In this framework, the network providing knowledge is called the teacher and the network learning the knowledge is called the student. During training, a student network learns to imitate the output of a larger and more powerful teacher network or ensemble of networks. Teacher-student learning frameworks have been widely used for performance improvement (especially for small datasets regimes) and/or model compression~\cite{hinton2015distilling}.\\

Inspired by the growing interest in applying machine learning to medicine and how to reuse and adapt previously acquired knowledge on new medical tasks and domains quickly, we propose adopting a teacher-student learning framework in the medical imaging setting. To the best of our knowledge, there is no study exploring a teacher-student learning framework to improve the performance of medical imaging diagnostic models. In this study, we conducted an empirical investigation to gather the advantages of knowledge transfer in medical imaging through a series of experiments. We focused on four main questions that we found to be fundamental in deriving our experimental analysis in the context of medical imaging:

\begin{itemize}
    \item How does knowledge transfer perform on small datasets?
    \item How does knowledge transfer perform when the domains and tasks are distinct?
    \item How much training data is needed to achieve high performance in knowledge transfer?
    \item Does knowledge transfer help with overfitting in a small data regime?
\end{itemize}

In terms of the teacher-student learning framework, we leveraged the work proposed in~\cite{ZagoruykoK16a} where the knowledge transfer is framed as an attention transfer mechanism. More specifically, a teacher network improves the performance of another student network by providing information about where it looks, i.e., about where it concentrates its attention.\\

Our experiments were conducted on two medical imaging diagnostic tasks: (1) Chest X-ray pathology classification and
(2) Diabetic Retinopathy (DR) classification.
%Radiographic imaging
The former, such as chest X-ray imaging is widely used in diagnosing several diseases such as thorax disease~\cite{guan_diagnose_2018}, Tuberculosis~\cite{lakhani_deep_2017}, Pneumonia~\cite{rajpurkar_chexnet_2017}, and COVID-19~\cite{cohen2020predicting}. Staff shortage in radiology departments in several countries~\cite{noauthor_clinical_2017, rosenkrantz_county-level_2018, nishie_current_2015} may put the patients life at risk.  This problem is even more severe 
in some countries such as Rwanda where there is one radiologist per 1000 patients~\cite{rosman_imaging_2015}, or in Liberia, 1 radiologist per 2 million patients~\cite{shortage2015}. The later, Diabetic Retinopathy, is also one of the major causes of blindness in the western world~\cite{pratt2016convolutional}. The early diagnosis of DR is crucial for its treatment. Early identification and scaling of DR involve localizing and weighting of numerous features on the Retina images which are highly time consuming. Both applications could benefit from recent advances in DNN and computer vision.\\

This paper is organized as follows: Section II summarizes related works. Section III describes the datasets used in this study. Section IV presents our proposed approach in building knowledge transfer including transfer learning and teacher-student framework. Section V presents our experiments and results. Section VI discusses the takeaways, addresses limitations of the current work, and proposes potential future work. 

\section{Related works}
\subsection{Medical imaging diagnostic}
\textbf{Chest X-ray pathology classification.} Enriched with access to the large public hospital scale datasets~\cite{wang_chestx-ray8:_2017, johnson_mimic-cxr:_2019, irvin_chexpert:_2019, bustos_padchest:_2019}, CNNs have been utilized for abnormality classification on medical chest X-rays images~\cite{irvin_chexpert:_2019, rajpurkar_chexnet_2017,rajpurkar_deep_2018,wang_chestx-ray8:_2017, CheXclusion_2020,yao_learning_2017}. The CNN classifiers are built to yield the diagnostic labels where the networks are trained on chest X-ray images and produce the probability of several diagnostic diseases per image. Transfer learning has been widely adopted for chest X-ray diagnostic tools~\cite{irvin_chexpert:_2019, rajpurkar_chexnet_2017,rajpurkar_deep_2018,wang_chestx-ray8:_2017, CheXclusion_2020,yao_learning_2017} and DenseNet~\cite{huang_densely_2017} is commonly used in training classifiers~\cite{zech_confounding_2018,rajpurkar_chexnet_2017, rajpurkar_deep_2018, yao_learning_2017, CheXclusion_2020, irvin_chexpert:_2019}. In addition to DenseNet, Irvin \emph{et al.}~\cite{irvin_chexpert:_2019} has applied several other CNN models including ResNet-152, Inception-v4, and SE-ResNeXt-101 on X-ray images, however, DenseNet-121 architecture was found to produce the best results in practice.\\
\begin{table*}[h]
\caption{\small A summary of medical imaging datasets used in this study. }
\centering
\begin{center}

\begin{tabular}{lccccc}
\toprule
\bf Dataset 	& \bf \# Labels & \bf Labeling Method  & \bf Images view	& \bf \# Images	& \bf \# Patients\\ \midrule
% Chest-Xray8 \cite{wang_chestx-ray8:_2017}	 & Automatic   & Frontal          & 112,120   	& 30,805\\ \hline

MIMIC-CXR~\cite{johnson_mimic-cxr:_2019}  	& 14  &Automatic    & Frontal/Lateral         & 371,858  	& 65,079\\ 
CheXpert~\cite{irvin_chexpert:_2019}  	& 14  &Automatic    & Frontal/Lateral         & 223,648   	& 64,740\\ 
ChestX-ray8~\cite{wang_chestx-ray8:_2017}	 & 15 &Automatic   & Frontal          & 112,120   	& 30,805\\ 
Retina~\cite{Cuadros2009-retina}  	&  5 &Manual    & Left/Right eyes         &  35,126  	& 17,563\\
\bottomrule
\end{tabular}
\end{center}
\label{tbl:datasets}
\vspace{-20pt}
\end{table*}

\textbf{Diabetic Retinopathy classification.} There has been a great amount of research for early detection of DR using neural networks~\cite{gardner1996automatic,nayak2008automated} and CNN~\cite{pratt2016convolutional}. However, insufficient annotated Retina dataset remains to be one of the challenges of applying deep learning in classification and early detection of DR. Transfer learning, therefore, has been extensively used to improve the performance of the models~\cite{li2017convolutional,masood2017identification,lam2018automated,benson2018transfer,retina-lit-1-Gulshan2016}. Although the CNN model achieved high accuracy for the binary classification of the disease using transfer learning, the performance degraded with increasing in the number of classes. This happens due to the imbalanced nature of the annotated data for some specific classes~\cite{lam2018automated}. In a study conducted by Gulshan \emph{et al.}~\cite{retina-lit-1-Gulshan2016}, it was shown that CNN models achieved high sensitivity and specificity for detection of diabetic retinopathy from Retinal fundus photographs. Raghu \emph{et al.}~\cite{raghu_transfusion:_2019} also conducted experimental evaluations of deep and light CNN models with different initialization strategies for detection of diabetic retinopathy.

\subsection{Knowledge Transfer}
In order to tackle shortcomings with the basic transfer learning, several advanced approaches were proposed including Knowledge Distillation (KD) in the neural network which is a knowledge transfer between a teacher and a student network~\cite{hinton2015distilling}. The original idea behind the KD came from Bucilua {\em et al.}~\cite{bucilu2006model} where they proposed the idea of compressing the knowledge of a number of large ensemble base-level classifiers into a single smaller and faster model. This would reduce the computation and memory complexity of the models. This idea was later generalized by Hinton {\em et al.}~\cite{hinton2015distilling} in which a knowledge is transferred from a large DNN (teacher) to a small network (student) by minimizing the difference between the logits (the inputs to the final softmax) produced by the teacher model and those produced by the student model. Yim {\em et al.}~\cite{Yim_2017} proposed an approach that minimized the distance between the intermediate layers of the teacher and student networks. This method helps with faster optimization and better performance of the student network than a DNN trained from scratch. Moreover, using their approach, the student DNN can learn the distilled knowledge from a teacher DNN that is trained for a different task. Romero {\em et al.}~\cite{romero2014fitnets} also proposed another teacher-student framework, called FitNet, where they introduced intermediate-level hints from the teacher's hidden layers in addition to output layers to guide the training process of the student network. Using FitNet, the student network can learn an intermediate representation that is predictive of the intermediate representations of the teacher network. FitNet is able to train very deep student models with less parameters, which can generalize better and/or run faster than their teachers. Attention transfer proposed by Zagoruyko {\em et al.}~\cite{ZagoruykoK16a} is a teacher-student training scheme similar to FitNet for knowledge transfer using teacher’s feature maps to guide the learning of the student. Using this approach, given the spatial attention maps of a teacher network, the student network is trained to learn the exact behavior of the teacher network by trying to replicate its output at a layer receiving attention from the teacher. The number of attention transfer and position of the layers depend on whether low-, mid-, and high-level representation information is required.\\

Motivated by advances in knowledge transfer approaches and their potential impact on medical image analysis, this study explores the performance of different training strategies in the context of transfer learning and teacher-student learning framework. This study's teacher-student learning framework is leveraging attention transfer mechanism for medical imaging diagnostic.

\section{Datasets}
\begin{figure*}[h]
    \centering
      \includegraphics[scale=0.5]{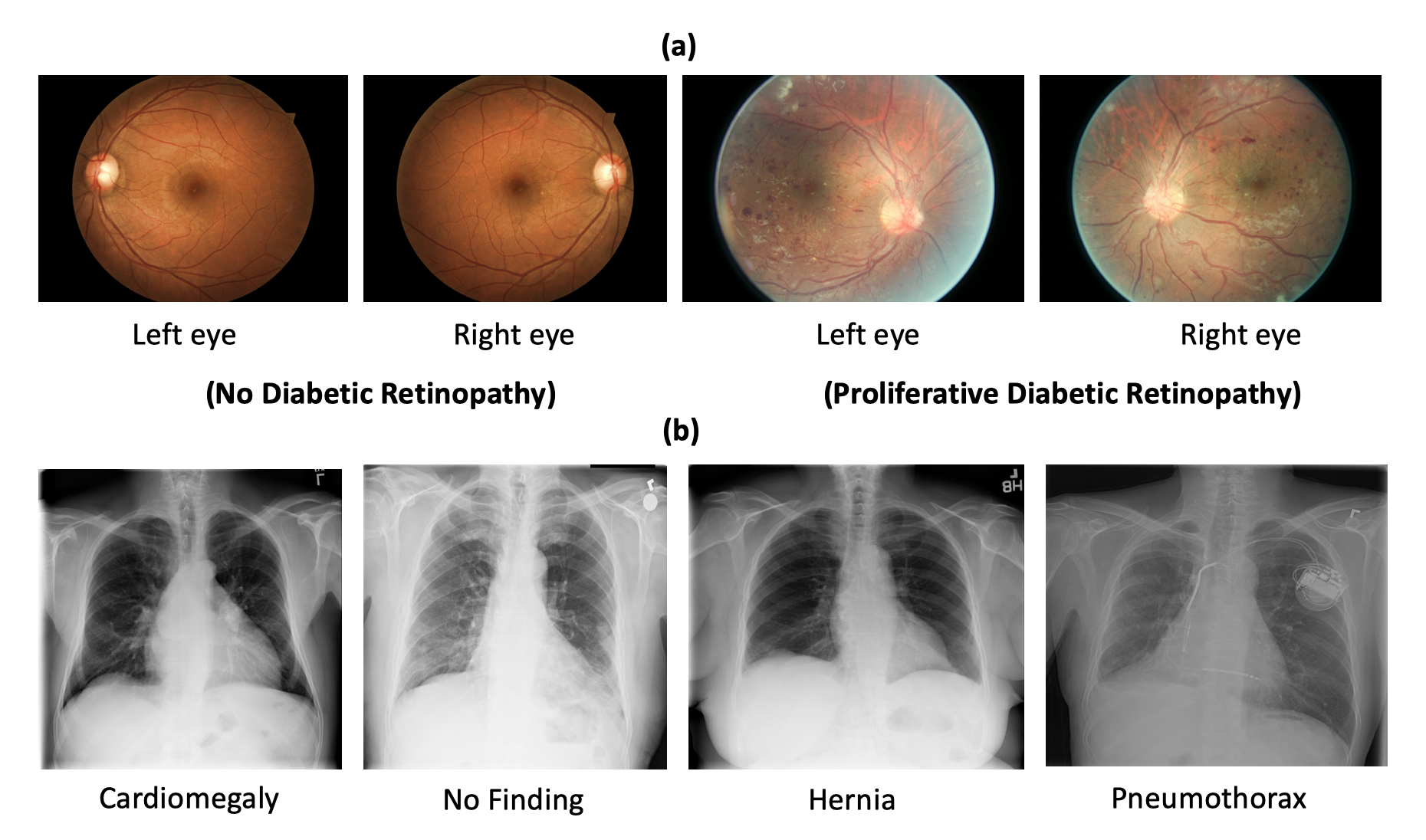}
      \caption{\small (a) Two examples of Diabetic Retinopathy (Retina) datasets for both right and left eyes. One example indicates no presence of Diabetic Retinopathy and the other example indicates presence of Profilerative Diabetic Retinopathy. (b) Four examples of the ChestX-ray8 dataset for three diseases (Cardiomegaly, Hernia, and Pneumothorax) and no finding.}
      \label{MedImgSample}
  \end{figure*}

In this study, we conducted our knowledge transfer experiments on four different publicly available medical imaging datasets listed in Table~\ref{tbl:datasets}. CheXpert~\cite{irvin_chexpert:_2019}, ChestX-ray8~\cite{wang_chestx-ray8:_2017}, and MIMIC-CXR~\cite{johnson_mimic-cxr:_2019} are chest X-ray images annotated for a number of diseases and Diabetic Retinopathy (Retina)\footnote{\url{https://www.kaggle.com/c/diabetic-retinopathy-detection}} is Retina images annotated for the diabetic scale of retinopathy. Fig.~\ref{MedImgSample} shows some sample images included in these datasets.\\

\textbf{CheXpert.} CheXpert~\cite{irvin_chexpert:_2019} is a chest radiographs dataset comprising 223,648 frontal and lateral images of 64,740 patients. Each image in the dataset has 14 multilabel annotations associated with diagnostic labels for 13 diseases: Enlarged Cardiomediastinum, Cardiomegaly, Lung Lesion, Lung Opacity, Edema, Consolidation, Pneumonia, Atelectasis, Pneumothorax, Pleural Effusion, Pleural Other, Fracture, Support Devices, and No Finding.\\

\textbf{ChestX-ray8.} The original ChestX-ray8~\cite{wang_chestx-ray8:_2017} includes 112,120 frontal X-ray images from 30,805 unique patients. However, in this study, we used a small sample (5\%) of the dataset translating to 5,606 images\footnote{\url{https://www.kaggle.com/nih-chest-xrays/data}}. ChestX-ray8 dataset includes 15 multiclass annotations for 14 diseases: Hernia, Pneumonia, Fibrosis, Edema, Emphysema, Cardiomegaly, Pleural Thickening, Consolidation, Pneumothorax, Mass, Nodule, Atelectasis, Effusion, Infiltration, and No Finding.\\

\textbf{MIMIC-CXR.} MIMIC-CXR~\cite{johnson_mimic-cxr:_2019} is a chest X-ray dataset composed of 371,858 frontal and lateral images of 65,079 patients. Similar to CheXpert, each image is annotated for the same 14 diagnostic diseases.\\

For all chest X-ray datasets (CheXpert, MIMIC-CXR, and ChestX-ray8), the labels were automatically extracted from the radiologist reports, using natural language processing techniques. For CheXpert and MIMIC-CXR in particular, the disease labels are from the set of \{positive, negative, not mention, or uncertain\} conditions. In this study, all "non-positive" labels were mapped to zero similar to “U-zero” study in~\cite{irvin_chexpert:_2019}. In all three chest X-ray datasets the "No Finding" label is not independent of the other disease labels and indicates absence of other diseases.\\

\textbf{Diabetic Retinopathy.} Retina dataset\footnote{\url{https://www.kaggle.com/c/diabetic-retinopathy-detection}} includes 35,126 high resolution RGB Retina images from both the right and left eyes of 17,563 patients. Each image is manually annotated by clinicians and represents the scale of presence of diabetic retinopathy: No DR, Mild, Moderate, Severe, and Proliferative DR. The images in the dataset are captured with different cameras affecting the visual appearance of left versus right eyes. Some images are inverted where the macula and optic nerve are flipped, as one sees in a typical live eye exam.

\section{Methods}
In the following section, we describe different knowledge transfer strategies conducted in this study to predict the diagnostic labels from medical imaging datasets. We focus on two knowledge transfer strategies: Transfer Learning and Teacher-Student Learning framework.

\subsection{Model Descriptions}
We used DenseNet~\cite{huang_densely_2017} as the backbone for our CNN classifiers. DenseNet is one of the latest neural networks for visual object recognition that has been used extensively in medical image classifications~\cite{rajpurkar_chexnet_2017, rajpurkar_deep_2018, irvin_chexpert:_2019,CheXclusion_2020,li2018thoracic}. DenseNet is composed of DenseBlocks and Transition Layers and the input to each layer of the DenseBlock is from all preceding layers. Transition Layers are placed between the Dense layers which includes batch normalization, a convolution layer and pooling layers to reduce the size and complexity of the model. For each task, depending on the dataset, we added an additional final layer to generate relevant predictions.
\\

Two versions of the DenseNet were used in this study; DenseNet-121~\cite{huang_densely_2017} and DenseNet-40. The latter is lighter than DenseNet-121 where we removed the last two blocks of the network for this study and we call it DenseNet-40 in the rest of the paper (see Appendix for more details of DenseNet-40). For the transfer learning approach, all DenseNet networks were initialized with ImageNet weights. For the teacher-student learning framework, the knowledge was transferred from a teacher either pre-trained on ImageNet (Teacher\textsubscript{ImageNet})~\cite{imagenet_cvpr09} or carefully pre-trained on MIMIC-CXR (Teacher\textsubscript{MIMIC-CXR})~\cite{johnson_mimic-cxr:_2019}. For the Teacher\textsubscript{MIMIC-CXR}, we leveraged PyTorch checkpoints provided by the work of Seyyed-Kalantari \emph{et al.}~\cite{CheXclusion_2020}. Teacher\textsubscript{MIMIC-CXR} is the DenseNet-121 initialized with the ImageNet and trained on 80\% of the MIMIC-CXR dataset. More details of the optimization and hyperparameter tuning of the network are reported in~\cite{CheXclusion_2020}.

\subsection{Attention Transfer}
\begin{figure*}[h!]
  \centering
  \includegraphics[scale=0.8]{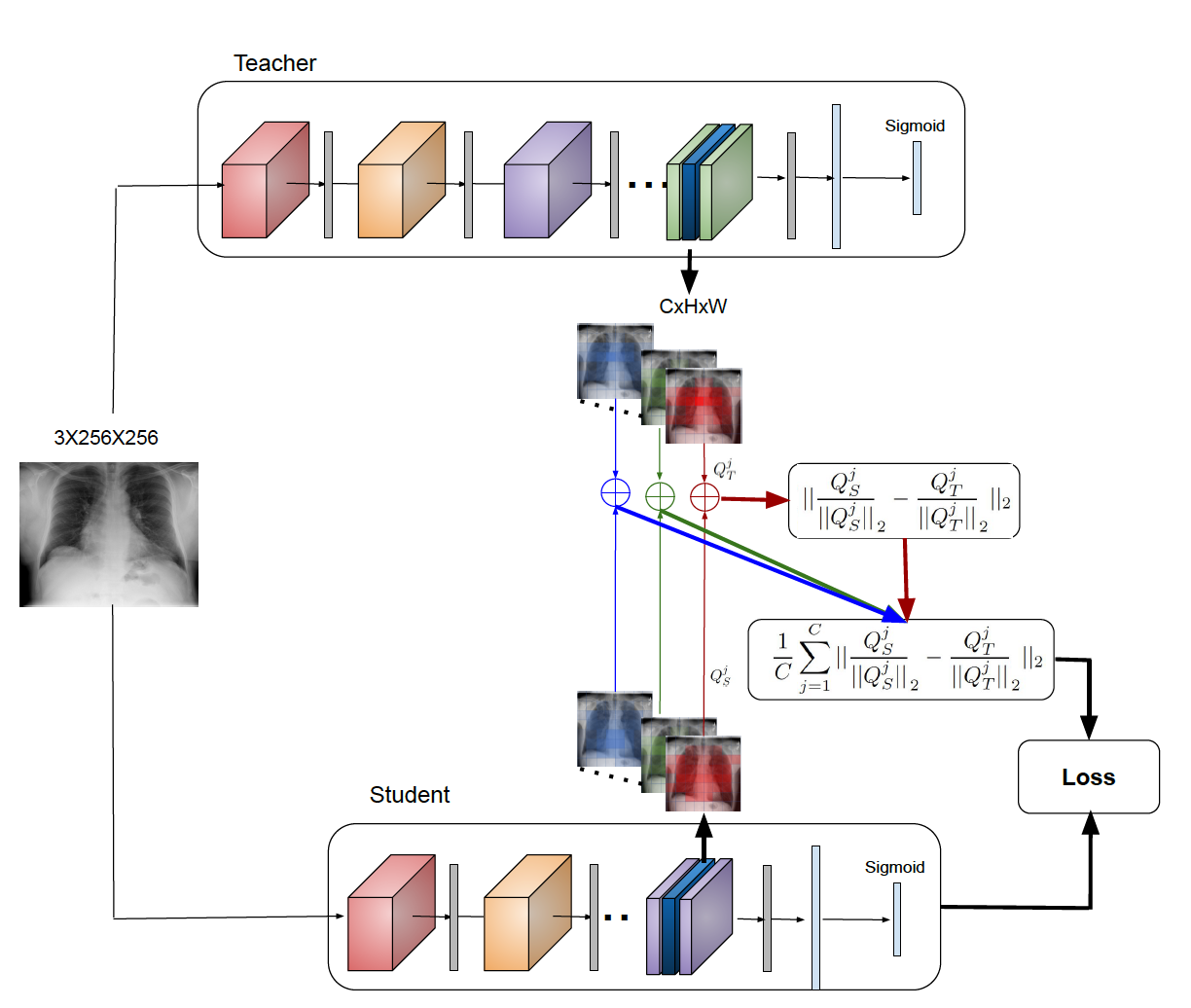}
  \caption{\small Illustration of Teacher-Student Learning framework using Attention Transfer map for knowledge transfer from a powerful CNN teacher to a CNN student. During training, the student network learns similar spatial attention maps to those of an already pre-trained teacher in order to make a good prediction. In our setting, transfer of knowledge occurs between the one layer before the last layer of the last dense blocks of both the teacher and student networks. In the shown example, the spatial attention map ($H \times W$) is $8\times 8$ and there are 32 feature planes ($C$).}
\label{scheme}
\end{figure*}

Following the work of Zagoruyko \emph{et al.}~\cite{ZagoruykoK16a}, we built an activation based attention transfer to transfer knowledge from a convolutional layer of the teacher network to a convolutional layer of the student network. In our setting, the knowledge was transferred between the one layer before the last layer of the last dense blocks of both the teacher and student networks as shown in Fig.~\ref{scheme}.\\

For a given convolutional layer, the corresponding 3D activation tensor, $A\in R^{C\times H\times W}$, consists of $C$ feature planes with spatial dimensions $H\times W$. We assume that transfer loss is placed between student and teacher attention maps with the same spatial resolution (same $H$ and $W$) as defined below:

\begin{equation}
L_{AT}=\frac{1}{C}\sum\limits_{j=1}^{C}   ||{\frac{Q_S^j}{||Q_S^j||}_2 -\frac{Q_T^j}{||Q_T^j||}_2}||_2,
\label{eq:at}
\end{equation}

where $Q_T^j$ and $Q_S^j$ are respectively the j-th feature plane (out of $C$ feature planes) of teacher's and student's 3D activation tensor, $A$, in a vectorized form. In order to calculate attention transfer loss, $Q_T^j$ and $Q_S^j$ were replaced with their $l_2$ normalized form as can be seen in Eq.~\ref{eq:at} and illustrated in Fig.~\ref{scheme}. The attention transfer loss was calculated by making use of the $l_2$ norm between student's and teacher's normalized feature planes averaged over all feature planes, $C$.

The total loss was defined as follows:

\begin{equation}
L_{tot}=  {CE}_S+ \frac{1}{\beta}L_{AT},
\label{eq:lss}
\end{equation}

where ${CE}_S$ is the standard cross-entropy loss for the student network and $\beta$ is the weight balancing attention loss and cross-entropy loss. %a coefficient weighing the contribution of attention transfer loss to the total loss. 
In this study, the ${CE}_S$ is a multi-label binary cross-entropy for X-ray datasets and multi-class cross-entropy for the Retina dataset. 

\section{Experiments}
In the following section, we describe the setup and results of a series of experiments we ran concerning transfer learning and attention transfer mechanisms on various medical imaging datasets.

\subsection{Set-up}
\emph{\textbf{Parameters}}. Adam~\cite{adam} was used to optimize the loss function in all of the tasks. The learning rate was decreased by a factor of 2 over every 16 epochs from an initial value of $5\times10^{-5}$ as suggested in~\cite{CheXclusion_2020}. For all the experiments, the CNN models were trained for a maximum of 128 epochs with a batch size of 32. So that each batch could fit in Nvidia Titan XP 12 \textit{GB} GPU used for training the CNN models. All evaluations were made based on three repetitions of each model. The best model and sets of hyperparameters were chosen based on the best AUC performance on the validation sets across all epochs. In order to find an optimal value for the $\beta$ coefficient shown in Table~\ref{tbl:hyper}, we performed a grid search in the range of values from 1 to 2000 on the validation set. The $\beta$ coefficient is reversely related to total loss - that is, when the $\beta$ decreases, the impact of attention loss on the total loss increases.\\

\emph{\textbf{Architecture}}. For both transfer learning and attention transfer settings, all 40 layers of DenseNet-40 ($1.4m$ trainable parameters) were unfrozen so their weights could get updated in each epoch of training. For the DenseNet-121 in the transfer learning setting, we conducted two tests where all the 121 layers ($7.0m$ trainable parameters) and the last 34 layers ($2.4m$ trainable parameters) of the network were unfrozen, respectively. For the attention transfer setting, both ImageNet and MIMIC-CXR were used for training the teacher network to explore the effect of cross-domain and in-domain training. In both cases, the teacher-student networks were initialized using ImageNet weights. Thus, the same initialization (i.e., ImageNet) was used for transfer learning. In the transfer learning setting, the loss function was the multi-label binary cross-entropy for X-ray datasets and multi-class cross-entropy for the Retina dataset. In the attention transfer framework, the cross-entropy loss was combined with attention transfer loss shown in Eq.~(\ref{eq:lss}). For the CheXpert and ChestX-ray8 datasets, the best models were selected based on the performance of the average of multi-label AUC on the validation set. For the Retina dataset, the best model was selected based on the weighted average of F1-score on the validation set. \\

\emph{\textbf{Data Augmentation}}. All of the images were resized to 256$\times$256, center cropped. Additionally, $-15^{\circ}$ to $+15^{\circ}$ random rotation and random horizontal flip were applied on the training dataset. Following~\cite{irvin_chexpert:_2019, rajpurkar_deep_2018, rajpurkar_chexnet_2017}, images were normalized using the mean and standard deviation of the ImageNet. All datasets were split into the train-validation-test-set split as listed in Table~\ref{tbl:datasetdivision} with no patient shared across the splits.

\begin{table}[h]
\caption{\small All medical imaging datasets were split into train-validation-test-set split with no patient shared across splits.}
\centering
\begin{center}
\begin{tabular}{lcccc}
\toprule
\bf Dataset & \bf Train & \bf Validation& \bf Test \\
\hline
CheXpert\textsubscript{5\emph{k}}  & 5,000 & 23,022 & 22,274\\
CheXpert\textsubscript{50\emph{k}} & 50,000 & 23,022 & 22,274\\
CheXpert\textsubscript{220\emph{k}} & 178,352 & 23,022 & 22,274\\
ChestX-ray8\textsubscript{5.6\emph{k}} & 3,589 & 848& 1,169\\
Retina & 21,126 & 7,000&7,000 \\
\bottomrule
\end{tabular}
\end{center}

\label{tbl:datasetdivision}
\vspace{-10pt}
\end{table}

\begin{table}[t]
\centering
\caption{\small Optimal values of $\beta$ used in attention transfer experiments. Optimal values were chosen through grid search between 1 and 2000 based on the the highest AUC on the validation set.}

\begin{center}
\begin{tabular}{lccc}
\toprule
\multirow{2}{*}{\bf Dataset}& \multirow{2}{*}{\bf Student}& \multicolumn{2}{c}{\bf Teacher} \\
\cmidrule{3-4}
&& MIMIC-CXR& ImageNet\\
\midrule
CheXpert\textsubscript{5\emph{k}}& DenseNet-121 & 1 & 2000 \\
CheXpert\textsubscript{5\emph{k}}& DenseNet-40 & 20 & 2000 \\
CheXpert\textsubscript{50\emph{k}}& DenseNet-121/DenseNet-40 & 50 & 2000\\
CheXpert\textsubscript{220\emph{k}}& DenseNet-121/DenseNet-40 & 1 &1000\\
ChestX-ray8\textsubscript{5.6\emph{k}}& DenseNet-121/DenseNet-40 & 1 & 100\\
Retina & DenseNet-121/DenseNet-40& 30&60 \\
\bottomrule
\end{tabular}
\end{center}

\label{tbl:hyper}
\vspace{-10pt}
\end{table}

\iffalse
\begin{table}[t]
\centering
\caption{\small Parameters chose for training attention transfer.}
\begin{center}
\begin{tabular}{lcccc}
\toprule
Dataset & Model & Teacher&$\beta$  \\
\hline
CheXpert\textsubscript{5\emph{k}}  & Large and Small & MIMIC&1 \\
CheXpert\textsubscript{50\emph{k}} & Large and Small & MIMIC& 50\\
CheXpert\textsubscript{xx\emph{k}} & Large and Small & MIMIC& 1\\
ChestX-ray8\textsubscript{xx\emph{k}} & Large and Small & MIMIC& 1\\
Retina & Large and Small & MIMIC&30 \\
CheXpert\textsubscript{5\emph{k}}  & Large and Small & Imagenet & 2000\\
CheXpert\textsubscript{50\emph{k}} & Large and Small & Imagenet & 2000\\
CheXpert\textsubscript{xx\emph{k}} & Large and Small & Imagenet &  1000\\
NIH\textsubscript{xx\emph{k}} & Large and Small & Imagenet & 100\\
Retina & Large and Small & Imagenet &60 \\

\bottomrule
\end{tabular}
\end{center}
\label{tbl:hyper2}
\vspace{-10pt}
\end{table}
\fi

\section{Results}

\emph{\textbf{1. Knowledge transfer for small datasets}}. Table~\ref{tbl:exp1} shows the performance of the knowledge transfer on small datasets of X-ray images. For this experiment, we trained our CNN models on a small subset of CheXpert\textsubscript{5\emph{k}} and ChestX-ray8\textsubscript{5.6\emph{k}} which were randomly sampled. At a high-level we observe that the teacher network pre-trained on MIMIC-CXR substantially improves the performance on both CheXpert\textsubscript{5\emph{k}} (AUC = 76.6$\pm$0.03) and ChestX-ray8\textsubscript{5.6\emph{k}} (AUC = 80.45$\pm$0.38) datasets. In this setting, student networks learn required knowledge for X-ray diagnostic tasks from a teacher pre-trained on chest X-ray imaging datasets. Hence, in a small dataset regime, attention transfer would improve the performance when the teacher and student networks are trained to learn the same task within a similar domain. However, for both datasets, a larger student network (DenseNet-121) with 6,968,206 trainable parameters outperforms a lighter student network (DenseNet-40) with 1,364,142 trainable parameters for attention transfer from Teacher\textsubscript{MIMIC-CXR}. For the CheXpert\textsubscript{5\emph{k}}, in particular, the AUC difference between large and light CNN models is very small which is not surprising as both teacher and student networks are trained on the similar domain which is X-ray images and same task which is classification for the same labels. On the contrary, the ChestX-ray8 has different sets of labels (diseases) compared with MIMIC-CXR but both are still in the same domain and the larger student network is significantly better than the lighter student network. Therefore, regardless of the domain of the source and target datasets, the student network should be deep enough to learn the task when the knowledge is transferred from a teacher pre-trained on a different task (different sets of labels). However, we emphasize the importance of utilizing attention transfer in order to improve the classification performance on the ChestX-ray8\textsubscript{5.6\emph{k}}. Using this approach teacher\textsubscript{MIMIC-CXR} provides an extra source of information where the CNN cannot gain if trained through a transfer learning approach as shown in Table~\ref{tbl:exp1}.\\

\begin{table*}[t]
\caption{\small The area under the receiver operating characteristic curve (AUC) score $\pm$ the 95\% confidence intervals (CI). The best scores are in bold, and the second best scores are underlined. *tp denotes trainable parameters. Here St~\textsubscript{DenseNet-40} and St~\textsubscript{DenseNet-121} denote the student network which is DenseNet-40 and DenseNet-121, respectively. For small imaging dataset, attention transfer improves the performance when knowledge is transferred from a teacher to a student trained for the same task (multi-label binary classification) within the same domain (chest X-ray).}
\centering
\begin{center}
\scalebox{0.9}{
\begin{tabular}{lcccccccc}%{@{}l|ccc|cccc@{}}
\toprule
\multirow{4}{*}{\bf Dataset} &&\multicolumn{3}{c}{\bf Transfer Learning} & \multicolumn{4}{c}{\bf Attention Transfer}\\
\cmidrule{2-9}
                && \multicolumn{2}{c}{\multirow{2}{*}{DenseNet-121}} & \multirow{2}{*}{DenseNet-40}& \multicolumn{2}{c}{Teacher~\textsubscript{ImageNet}}& \multicolumn{2}{c}{Teacher~\textsubscript{MIMIC-CXR}}\\
\cmidrule{6-9}
	& &  &  & 	& St~\textsubscript{DenseNet-121}	& St~\textsubscript{DenseNet-40} & St~\textsubscript{DenseNet-121}&  St~\textsubscript{DenseNet-40}\\
 \cmidrule{2-9}
 &tp* & 7.0\emph{m} & 2.7\emph{m} & 1.4\emph{m}& 7.0\emph{m} & 1.4\emph{m}& 7.0\emph{m} & 1.4\emph{m}\\
\midrule
CheXpert\textsubscript{5\emph{k}} & &72.86$\pm$0.18 & 71.23$\pm$0.12 & 71.83$\pm$0.23 & 72.64$\pm$0.12& 72.22$\pm$0.10 & \bf 76.6$\pm$0.03& \underline{75.45$\pm$0.30}\\ 
ChestX-ray8\textsubscript{5.6\emph{k}}& &71.45$\pm$ 0.98 & 70.07 $\pm$ 0.60 & 71.55$\pm$ 0.19 & 71.66$\pm$ 1.18 & 72.45 $\pm$ 1.09 & \bf 80.45$\pm$ 0.38 & \underline{75.79$\pm$ 1.17}\\
\bottomrule
\end{tabular}}
\end{center}
\label{tbl:exp1}
\vspace{-10pt}
\end{table*}

\emph{\textbf{2. Knowledge transfer between distinct domains and tasks}}. In this experiment, we trained CNN models on the Retina dataset which is different from both ImageNet and MIMIC-CXR. Some of the differences are as follows: (1) Retina and ImageNet images are RGB versus X-ray images which are grayscale. (2) all datasets have different sets of classes and labels, i.e., Retina and ImageNet are not multi-label and each image associated with only one of the 5 and 1000 class labels, respectively; however, X-ray has 15 multi-labels binary classes where one image may have more than one disease label positive. Taking the differences into account, it can be said that the Retina dataset is much closer to ImageNet than X-ray images. Table~\ref{tbl:exp2} shows the performances of the CNN models on the test set. Our results indicate that better AUC (85.04$\pm$0.28) and F1-score (78.01$\pm$0.55) performance are achieved on Retina images if the knowledge (attention in our case) is transferred from the teacher network pre-trained on ImageNet than MIMIC-CXR. These results imply that in a teacher-student learning framework, the performance substantially increases when the teacher network is pre-trained on a domain similar to the domain that student network will be trained on.\\

\begin{table*}[t]
\caption{\small The AUC and F1-scores $\pm$ the 95\% CI for Retina $(\approx35k)$ dataset. The best scores are in bold, and the second best scores are underlined. *tp here denotes trainable parameters. In the teacher-student learning framework, the performance of the student network is substantially increases when the teacher network is pre-trained on a domain similar to the student's domain.}
\centering
\begin{center}
\scalebox{0.9}{
\begin{tabular}{@{}lcccccccc@{}}
\toprule
\multirow{4}{*}{\bf Metrics} &&\multicolumn{3}{c}{\bf Transfer Learning} & \multicolumn{4}{c}{\bf Attention Transfer}\\
\cmidrule{2-9}
&& \multicolumn{2}{c}{\multirow{2}{*}{DenseNet-121}} & \multirow{2}{*}{DenseNet-40}& \multicolumn{2}{c}{Teacher~\textsubscript{ImageNet}}& \multicolumn{2}{c}{Teacher~\textsubscript{MIMIC-CXR}}\\
\cmidrule{6-9}
	&&   &  & 	& St~\textsubscript{DenseNet-121}	& St~\textsubscript{DenseNet-40} & St~\textsubscript{DenseNet-121}&  St~\textsubscript{DenseNet-40}\\
% \cmidrule{2-9}
%\cmidrule{2-8}
%                & \multicolumn{2}{c}{DenseNet121} & DenseNetxxx& \multicolumn{2}{c}{Teacher \textsubscript{ImageNet}}& \multicolumn{2}{c}{Teacher \textsubscript{Mimic}}\\
%\cmidrule{2-8}
%	& All blocks  & Last block & All blocks	& student \textsubscript{light}	& student \textsubscript{standard} & student \textsubscript{light}&  student \textsubscript{standard}\\
% & & & & & & & \\
 &tp* & 7.0\emph{m} & 2.7\emph{m} & 1.4\emph{m}& 7.0\emph{m} & 1.4\emph{m}& 7.0\emph{m} & 1.4\emph{m}\\
\midrule
AUC &&84.46$\pm$0.52 & 79.98$\pm$ 0.64 & 84.95$\pm$0.28 & \underline {84.61$\pm$0.19} & \bf 85.04 $\pm$0.28 & 84.03$\pm$0.31 & 84.38$\pm$0.33 \\
F1-score &&77.9$\pm$ 0.22 & 74.02 $\pm$ 0.19 & 77.73$\pm$ 0.44 & \underline {78.02$\pm$ 0.31} & \bf 78.01 $\pm$ 0.55 & 77.68$\pm$ 0.22  &77.93$\pm$ 0.17 \\
\bottomrule
\end{tabular}}
\end{center}

\label{tbl:exp2}
\vspace{-10pt}
\end{table*}

\begin{figure*}
  \centering
  \includegraphics[scale=0.37]{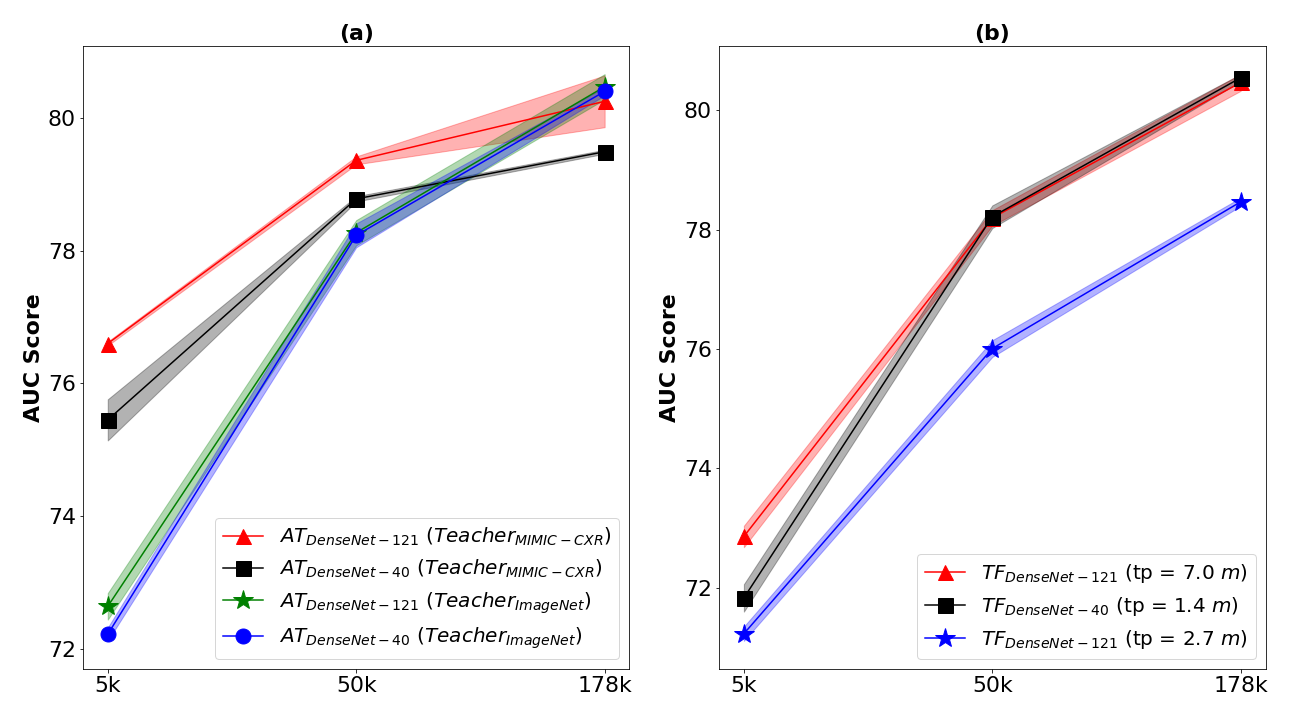}
  \caption{\small AUC scores of (a) attention transfer and (b) transfer learning as a function of the number of training examples sampled from CheXpert. Here \textbf{AT\textsubscript{\emph{DenseNet-121}} (Teacher\textsubscript{\emph{MIMIC-CXR}})} denotes that knowledge is transferred from Teacher\textsubscript{MIMIC-CXR} to the student network which is DenseNet-121. \textbf{TF\textsubscript{\emph{DenseNet121}} (tp = $7.0$ \emph{m})} denotes transfer learning using DenseNet-121 with $7.0m$ trainable parameters. The performance on AUC score for both attention transfer and transfer learning increases as the number of training data increase. For attention transfer, the AUC score difference between in-domain and cross-domain knowledge transfer decreases as the size of training data increases.
  }
\label{fig:exp3}
\end{figure*}

\emph{\textbf{3. The effect of dataset size on Knowledge Transfer}}. In order to analyze how much training data is needed to achieve high performance on attention transfer, we show AUC curves of CNN models as a function of the number of training examples sampled from CheXpert (5\emph{k}, 50\emph{k}, and entire data which is 178\emph{k}) in Fig.~\ref{fig:exp3}. A glance at the plots reveals three trends. First, for both transfer learning and attention transfer, the performance on AUC score increases as the number of training data increases. Second, in-domain attention transfer substantially outperforms cross-domain attention transfer for a small training data but as the size of training data increases in-domain and cross-domain attention transfer perform the same. As it can be seen from Fig.~\ref{fig:exp3}-(b), the same is true for transfer learning setting as well. This can be explained by the fact that for large dataset the network can learn the task from the data through optimizing the cross-entropy loss, therefore there is less need to reuse previously acquired knowledge through attention transfer. Lastly, for the transfer learning approach from ImageNet, regardless of the size of medical imaging data, it is always beneficial to unfreeze all CNN layers during training so all parameters of the network get updated at each epoch.\\

\begin{figure*}[h!]
  \centering
  \includegraphics[scale=0.7]{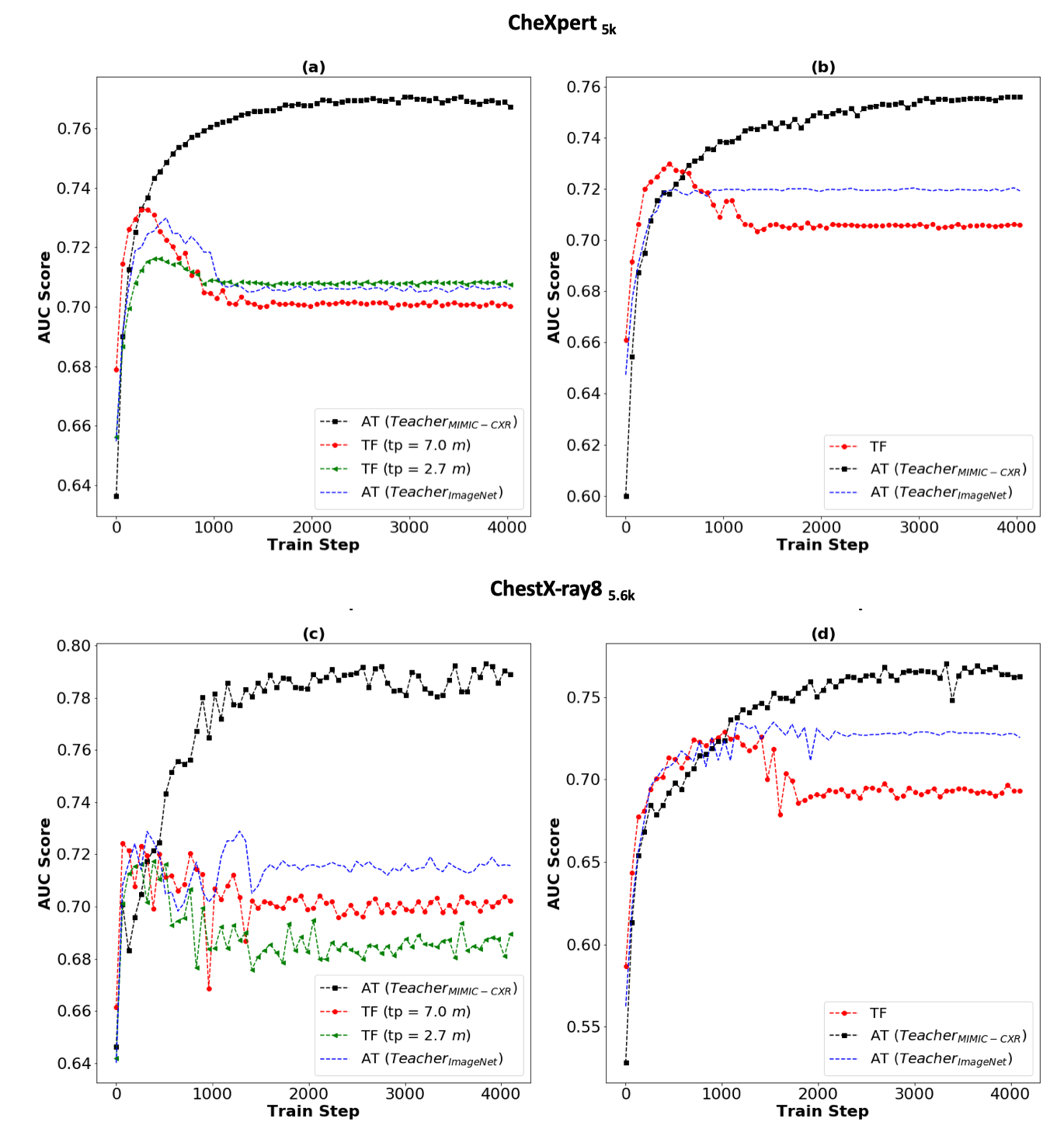}
  \caption{\small The vertical and horizontal axes indicate AUC score and training steps, respectively, for training CNN models on CheXpert\textsubscript{5\emph{k}} (top) and ChestX-ray8\textsubscript{5.6\emph{k}} (bottom). Student network is DenseNet-121 in (a) and (c), and DenseNet-40 in (b) and (d). Attention transfer (AT) as a regularizer delays the over fitting and makes the training more robust comparing with transfer learning (TF) approach. However, it might slows down the convergence but it allows the student network to continue training.}
\label{conv}
\end{figure*}

\emph{\textbf{4. Knowledge transfer as a regularizer.}} Fig.~\ref{conv} illustrates the AUC learning curves of CNN models trained using student-teacher framework on CheXpert\textsubscript{5\emph{k}} and ChestX-ray8\textsubscript{5.6\emph{k}} validation set. Our analysis indicates that regardless of the size of the student network (DenseNet-40 or DenseNet-121), attention transfer not only improves the performance but also serves as a regularizer to delay overfitting. The regularization effect of attention transfer stabilizes the training of the student network with less fluctuations.

\section{Discussion}
Knowledge transfer is widely used in computer vision tasks to enable deep CNN models to quickly learn complex concepts when trained on small image datasets (e.g., hundred/thousands versus millions of images). In this paper, we provided further insights into the adoption of the teacher-student learning framework based on the concept of attention transfer~\cite{ZagoruykoK16a} on training CNN models for the medical domain. Our experiments were conducted on  diagnostic classification tasks where we explored fundamental components of knowledge transfer on three medical imaging datasets.\\

Our series of experiments revealed that in a small data regime (less than 50\emph{k}), regardless of the source and target domains, attention transfer outperforms transfer learning approach. However, as the size of the dataset increases attention transfer and transfer learning perform the same function. In terms of transfer learning approach, our finding was in line with the previous study~\cite{raghu_transfusion:_2019} where lighter CNN models were shown to have relatively similar performance results to that of larger networks. We also found that in the teacher-student learning framework, the performance of the CNN model depends on the similarity between the source and target domain where with more similar domains, higher performance can be achieved. However, it is important to note that the source and target domains do not necessarily need to be the same, yet, the best performance will be achieved when the knowledge is transferred from a teacher pre-trained on a similar domain to target domain. For instance, in the case of Retinopathy diabetic classification, the best performance was achieved through knowledge transfer from ImageNet than medical X-ray; however, all three domains are distinct where Retina and ImageNet are more similar than Retina and X-ray. One other interesting aspect of attention transfer is its regularization effect which delays the over fitting and makes the training more robust compared with the transfer learning approach. Although it might slow down the convergence but it will allow the network to continue training which improves the performance; a trade-off between improvement and convergence speed. To sum up, attention transfer can help with performance improvement on small medical imaging dataset and cross-domain knowledge transfer from other domains to medical imaging domain, and last but not least serve as a regularizer during training of the CNN on medical imaging dataset.\\

As already mentioned in the above paragraph, in this study we showed that the teacher-student learning framework significantly helps the performance for small imaging datasets. Limited availability of annotated data is a major issue in medical imaging and usually restrains medical research particularly at the beginning of any pandemic. Currently, the 2019 novel coronavirus (COVID-19) is affecting the world and a small amount of available data hurts researchers’ ability to build machine learning models that can help inform decision-makers with a timely response to the disease. COVID-19 is a virus that directly affects the lungs, so chest X-ray or CT routinely used by clinicians for diagnosis of pneumonia~\cite{garin2019computed} have the potential to be leveraged for COVID-19 screening in emergency departments and ambulatory settings. As a result, there have been recent efforts in the machine learning community to develop advanced computer vision models for automated detection of COVID-19 cases from medical images~\cite{cohen2020predicting}. This is an example of the need for advanced knowledge transfer techniques that can improve performance of DNN diagnostic models to be trained on small datasets. 

\subsection{Limitations and Future directions}
Our study, being of an exploratory analysis, raises a number of opportunities for future work which would further elaborate knowledge transfer in the medical setting.

\emph{\textbf{Explainability:}} In this study, we highlighted some of the advantages of attention transfer versus transfer learning including a higher performance of the network for small datasets and acting as a regularizer during training. There is another important advantage of leveraging attention transfer in the medical settings which is its capability in providing some level of explainability and it has not been explored in this study. A direction for future research that stems from this work is to analyze the attention weights to explore where the medical image teacher and student network pay more attention to, and the potential correlation of attention weights with the disease. \\

\emph{\textbf{Knowledge distillation:}} Attention transfer is just one form of teacher-student learning framework. There are various forms of knowledge transfer that have been widely developed and implemented to aid generalization while training deep CNN models on various domains and tasks. A potential future work could be investigating other forms of knowledge transfer especially knowledge distillation models on medical imaging mainly for model compression that is more suitable for deployment on edge devices. \\

\emph{\textbf{Initialization strategy:}} During training and optimization of a CNN, the search space for the optimal parameters is determined by the choice of hyperparameters and the initialized weights of the network as well as the training strategy. One of the limitations of this study was to restrict our experiments to examine different training strategies of CNN models initialized with ImageNet weights only. In the attention transfer framework, we explored knowledge transfer for both in-domain and cross-domain settings where in both cases, the network initialization was the same as that of transfer learning setting (e.g., ImageNet). In transfer learning, we didn't study networks’ performances initialized with cross-domain pre-trained weights since the goal was to use the same initialization for both transfer learning and attention transfer. Exploring different initializations in order to expand the search space for optimization of CNN models is an interesting research direction to be pursued in future. \\

\emph{\textbf{Few-shot learning:}} The other direction that remains to be investigated is combining attention transfer with few-shot learning. Inspired by the work of Tian \emph{et al.}~\cite{53-tian}, a novel future research is to combine attention transfer with few-shot learning techniques to learn a good embedding that can generalize well on a novel class. In this setting, it is essential to learn a good embedding such that when we apply the model that is trained on a base dataset (e.g CheXpert), it can predict a novel class (e.g COVID-19), with access to very limited images from the novel class at test time only. In other words, no image of the novel class is offered to the network throughout training or/and validation. \\

\emph{\textbf{COVID-19 early detection:}} As mentioned above, our proposed attention transfer pipeline enhances the performance of CNN models in a small training data regime where access to a large annotated dataset is not possible. This was the case in the early stages of the COVID-19 pandemic where the large datasets of COVID-19 may have less than a thousand positive images~\cite{covid-laleh-1,covid-laleh-2}. Thus, a trained  
%In order to verify the usefulness of our proposed approach, we conducted a pilot study where we trained a%
 CNN using a teacher-student framework can be utilized for early detection of COVID-19 from chest X-ray or CT (computed tomography) images\footnote{\url{https://github.com/ieee8023/covid-chestxray-dataset}} with access to less amount of images. %Our preliminary results were encouraging enough to inspire the future of this work. 
 %In our future work, we plan to expand this idea to use %include more types of 
 %medical images such as CT to diagnosis of infectious diseases and COVID-19 in particular.

\section*{Acknowledgment}
We would like to acknowledge Vector Institute and also its high performance computing platforms made available for conducting the research reported in this paper. We also like to thank Vanessa Allen, Samir Patel, and Public health Ontario for their support in this project. We also acknowledge the support of the Natural Sciences and Engineering Research Council of Canada (NSERC), [funding reference number PDF-516984].
  \newpage
%\fi



%\bibliographystyle{IEEEtran}
\bibliographystyle{IEEEtran}  
\bibliography{references}

\newpage
\appendices
\section{}
\subsection{Details of Light DenseNet}\label{appndx}
In this study, we explored teacher-student learning frameworks in the medical imaging setting for both large and light convolutional neural network students. DenseNet-121 was used for the large student network. We built a light version of DenseNet, called DenseNet-40, for the light student network. In DenseNet-40, we removed the last two blocks of DenseNet-121. Details of DenseNet-40 architecture are shown in Table~\ref{tbl:arch-smal}.

\begin{table}[htb]
\centering
\caption{\small The architecture of light DenseNet called DenseNet-40. In DenseNet-40, we removed the last two blocks of DenseNet-121 for the purpose of our exploration.}

%\bgroup
\def\arraystretch{1.6}
\begin{center}
\scalebox{0.9}{
\begin{tabular}{lcl}
\hline
\bf Layers                                                    & \bf Output size &     \bf Filters                               \\ \Xhline{3\arrayrulewidth}
Convolution                                               & 128 $\times$ 128     & 7 $\times$ 7 conv, stride 2               \\\hline
Pooling                                                   & 64 $\times$ 64       & 3 $\times$ 3 max pool, stride 2           \\\hline\hline
Dense Block (1)                                           & 64 $\times$ 64       &  \(\begin{bmatrix} 1&\times&1&\text{conv}\\
3&\times&3&\text{conv}
\end{bmatrix}\) $\times$ 6                              \\ \hline
\multicolumn{1}{c}{\multirow{2}{*}{Transition Layer (1)}} & 64 $\times$ 64       & 1 $\times$ 1 conv                         \\
\multicolumn{1}{c}{}                                      & 32 $\times$ 32       & 2 $\times$ 2 average pool, stride 2       \\\hline\hline
Dense Block (2)                                           & 32 $\times$ 32       &   \(\begin{bmatrix} 1 &\times& 1&\text{conv}\\
3 &\times& 3&\text{conv}
\end{bmatrix}\) $\times$ 12                                 \\\hline\hline
Pooling                                                   & 8 $\times$ 8         & 8 $\times$ 8 Adaptive Average Pool        \\\hline
Convolution                                               & 8 $\times$ 8         & 1 $\times$ 1 conv                         \\\hline
Convolution                                               & 8 $\times$ 8         & 1 $\times$ 1 conv                         \\\hline
Pooling                                                      & 1 $\times$ 1         & 1 $\times$ 1 Adaptive Average Pool        \\\hline\hline
Classification Layer                                      &             & fully-connected, Softmax / Sigmoid \\\hline\hline

\end{tabular}
}

\end{center}

\label{tbl:arch-smal}
\vspace{-10pt}

\end{table}
%\begin{thebibliography}{1}


%\end{thebibliography}



%\newpage


%\vfill


\end{document}